\documentclass[referee]{aa}

\usepackage{graphicx}

\begin{document}
 
\title{Neutron stars and strong-field effects of general relativity}

\titlerunning{Neutron stars and general relativity}

\author{ W{\l}odek Klu\'zniak}

\institute{Copernicus Astronomical Center, ul. Bartycka 18, 00-716 Warszawa,
Poland}

\authorrunning{Klu\'zniak}

\maketitle

\begin{abstract}
The basic observed properties of neutron stars are reviewed.
I suggest that neutron stars in low-mass X-ray binaries
are the best of all known sites for testing strong-field effects of general
relativity.

\end{abstract}

\section{Validity of General Relativity}

General Relativity (GR) is the correct description of gravity and space-time.
The phenomena verified with three classic tests of GR are so well established
that they are now used as tools in every-day astronomical practice and
even in technological applications.

The gravitational bending of light, famously detected in Eddington's
solar eclipse expedition is today used to determine the stellar content of
our Galaxy and the Magellanic Clouds (from stellar micro-lensing events
detected by the OGLE, MACHO and EROS experiments). Lensing of distant
galaxies by intervening galaxy clusters is used to determine the (dark)
matter distribution in the latter.

Gravitational redshift, first observed in spectra of the white dwarf
Sirius B in 1925, has since been detected in the laboratory (Pound-Rebka
experiment) and is now of necessity taken into account in surveying
practice (the GPS system). The effect is also essential
in timing radio pulsars---when compared to some millisecond pulsars,
terrestrial clocks clearly run slower at full moon than at new moon.

The magnitude of precession of the perihelion of Mercury is dwarfed
by the same effect in the Hulse-Taylor pulsar, where the periastron
shifts by $4.2^\circ$ per year. A similar system, Wolszczan's
binary pulsar, allows a confirmation of the Shapiro delay.

Of course, GR also provides the framework for understanding the evolution
of our expanding Universe. All these successes allow us to confidently
use general relativity, even in domains where its validity has not yet
been strictly proven.

Observations of certain X-ray binaries (e.g., Cygnus X-1 and the so called
X-ray novae), as well as of stellar motions in our Galaxy, and of velocities
in the inner cores of other galaxies, strongly suggest the existence of
black holes. However, the laws of GR have not yet been truly tested in the
strong field regime.

\subsection{Why neutron stars}

The strength of gravity is conveniently parametrized by the mass to size
ratio, $(M/R)(G/c^2)$. For black holes, of course, $GM/(Rc^2)\sim1$,
as for the Schwarzschild radius $R_{\rm Sch}=2MG/c^2$. For the Sun,
$GM_\odot/c^2\approx 1.5\,$km, while the solar radius $R_\odot\approx300 000\,$
km, which yields $M_\odot/R_\odot\sim10^{-5}$ (in units of $c^2/G$).
A similar value is obtained for mass/distance in the
 binary Hulse-Taylor pulsar, where
relativistic effects in the orbital motion are so clearly detected
(because the pulsar period is so short $\approx0.06\,$s, and known
to 10 significant figures). For white dwarfs, $M/R\sim10^{-3}$.
But for neutron stars, $M/R\sim 10^{-1}$, and GR effects just outside their
surface are about as important as near the black hole surface.

As a testbed for GR, neutron stars have one great advantage over black
holes---they have a tangible surface which can support magnetic fields
and can emit X-rays and other radiation. A great deal can be learned 
about neutron stars without assuming the validity of GR.
Hence, a great deal can be learned about GR by observing neutron stars.
Today, about 1000 radio pulsars are known and about 100 X-ray binaries
containing neutron stars, so also in sheer numbers neutron stars
have an advantage over black holes.

\subsection{Basic references}

The narrative presented in Sections 1 and 2, to a large extent
relies on well established observations and theories, which have made
their way into excellent textbooks, where detailed references can be
found to the literature. Among those, particularly useful in the context
of these lectures are the ones by Shapiro and Teukolsky (1983),
Lipunov (1992), M\'esz\'aros (1992),
Glendenning (1997), and Frank, King and Raine (1985).

\section{A brief history of neutron stars}.

Before discussing in detail the properties of rapidly rotating,
(at most) weakly magnetized, compact stars---which are ideal astrophysical
objects for testing strong-field predictions of General Relativity---let us
recount how they were identified.

\subsection{Key dates}

The basic chronology of the discovery of neutron stars can be found,
together with the references, e.g., in the text by Shapiro and Teukolsky.
The following selection reflects my bias of what seems particularly
important with the hindsight of today.

1914: Adams discovered that the rather dim, $L\approx3\times10^{-3}L_\odot$,
star Sirius B (orbiting Sirius), whose mass had been determined to be
$M\approx0.85\pm0.10M_\odot$, has the spectrum of a ``white'' star---hence
the name white dwarf. The unusual combination of low luminosity and
high temperature implied a small radius, $R\approx2\times10^{4}\,$km.
This conclusion was based on an application of the black-body formula
$$L=4\pi\sigma_BR^2T^4. \eqno(1)$$

1925: Adams measures the redshift, $z$, of certain lines in Sirius B.
Applying general relativity, one can infer the value of $M/R$ from $z$,
and from the known mass a value of the stellar radius, $R\sim10^4\,$km.
The agreement with the spectroscopically determined value was a great
triumph of GR.

1926: The Fermi-Dirac statistic is discovered.

1926 (December): Fowler identifies the agent holding up white dwarfs
against gravity---it is the degeneracy pressure of electrons.

1930: Chandrasekhar discovers theoretical models of white dwarfs, from which
the maximum value for white dwarf mass follows, the famous $1.4M_\odot$.
Incidentally, $M\sim1M_\odot$ and $R\sim {\rm few} \times 10^3\,$km imply
a density $\rho\sim10^6\,$g/cm$^3$, which in turn implies a minimum
period of possible rotation or vibration of a few seconds:
$({G\rho})^{-1/2}\sim3\,$s.

1932: Chadwick discovers the neutron.

1932: Landau discusses cold, degenerate stars composed of neutrons.

1934: Baade and Zwicky write: ``With all reserve we advance the view that
supernovae represent the transition from ordinary star to neutron stars.''
This remains a  remarkable contribution---two years after the discovery of
neutrons, Baade and Zwicky correctly explain the mechanism of Supernovae
(type II) explosions, find the correct value for the gravitational
binding energy released in the creation of a neutron star, $\sim10^{53}\,$erg,
and even identify a site where a neutron star is present (and was discovered
35 years later!): the Crab nebula.

1938: Landau discusses the energy released
inside ordinary stars with neutron-star cores (a theoretical precursor
of what is now known as a Thorne-\.Zytkow object). At the time, the energy
source of the Sun was not known. The great contribution here is the
pointing out of the enormous energy released in accretion onto neutron
stars.

1939: Oppenheimer and Volkoff solve the relativistic equations of
stellar structure for a fermi gas of neutrons, and thus construct the
first detailed model of a neutron star. They find a maximum mass
($\approx0.7M_\odot$, lower than the one for modern equations of state),
above which the star is unstable to collapse. Thus the road to the
theoretical discovery of black holes is paved.

1940's are lost to the Second World War.

1950's: The basic physics of the interior of neutron stars is worked
out by the Soviet school, including a detailed understanding of the
superfluid phase.

1962: Giacconi et al. discover the first extrasolar source of X-rays,
Sco X-1.

1967: Shklovsky derives a model for Sco X-1, in which the X-ray source
is an accreting neutron star in a binary system.

1967: Pacini points out that neutron stars should rotate with periods
$P<<1\,$s, and may have magnetic fields of surface value $B\sim10^{12}\,$G.
The ensuing dipole radiation is not directly observable, as its frequency
$2\pi/P$ is below the plasma frequency of interstellar space.

1967: Radio pulsars with $P\le3\,$s discovered by Hewish, Bell et al.

1968: Gold gives the ``lighthouse'' model of radio pulsars.

1968: Spin-down of radio pulsars is measured, $\dot P>0$. From this
moment, it is clear that pulsars are rotating, compact objects, ultimately
powered by the kinetic energy of their rotation.

1971: Giacconi et al. discover the first of accreting counterparts of
radio pulsars, the X-ray pulsar Cen X-3, of period 4.84 s. Today, many
are known, in the period range $0.7\,{\rm s}\le P\le 10000\,$s.

1978: Tr\"umper et al. discover the $\sim 40\,$keV cyclotron line in
the spectrum of the accreting X-ray pulsar Her X-1. From the formula
$h\nu=1\,{\rm keV}\times(B/10^8\,{\rm G}$), the inferred value of the
magnetic field at the stellar surface is $B_p=\,$few$\times10^{12}\,$G,
in agreement with the estimates of the dipole strength of ordinary
radio pulsars.

1982: The discovery of millisecond pulsars by Backer, Kulkarni et al.

1996: The discovery of kHz quasi-periodic oscillations (QPOs) in
the X-ray flux of low-mass X-ray binaries (LMXBs).

1998: The discovery of 2.5 ms pulsar in the transient LMXB
SAX J 1808.4-3658 by Wijnands and van der Klis.

\subsection{The physics of identifying neutron stars}

It should be apparent from the above review, that the basic physics behind
identifying neutron stars is fairly simple. Of course, the discovery was
possible only after decades of sustained technological development,
particularly in the field of radio and X-ray detectors, as well as much
observational effort. Also, the existence of neutron stars would not have been
so readily accepted without the solid theoretical foundations laid down
over a period of many years. But the basic, incontrovertible,
observational arguments are really based on two or three simple formulae.

Let us accept the theoretical result, that a neutron star is a body of
mass $M\sim1M_\odot$ and radius $R\sim10\,$km, hence of mean density
$\bar \rho>10^{14}\,$g/cm$^3$. How can we be certain that such bodies have been
discovered?

a) The mass can be determined directly in some binary systems
by methods of classical astronomy (as developed for spectroscopic binaries),
essentially by an application of Kepler's laws.
For the binary X-ray pulsars, the errors are rather large, but it is clear
that one or two solar masses is the right value. For the binary radio
pulsars (the Hulse-Taylor and Wolszczan pulsars), where the pulse
phase can be determined very precisely and relativistic
effects give much redundancy, the mass has been measured very accurately
(to $0.01M\odot$) and is close to $1.4M_\odot$. For binary (millisecond) radio
pulsars with white dwarf companions, the mass function is always consistent
with these values.

b) In bright, steady, X-ray sources, and especially in X-ray bursters
(where the X-ray flux briefly saturates at a certain peak value),
one can assume that the radiative flux is limited, at the so called
Eddington value, by a balance between
radiation pressure on electrons and gravitational pull on protons.
Since both forces are proportional to (distance)$^{-2}$, there
is a direct relation between flux and mass. Again, $M\sim 1M_\odot$ is
obtained, for $L_X\approx10^{38}\,$erg/s.

c) The radius can be determined whenever a thermal spectrum is detected,
by a combination of the black-body formula, eq. (1), and of Wien's law
giving the characteristic temperature of a body emitting the thermal
spectrum. Thus, for X-ray pulsars, such as Her X-1, the spectrum gives
a characteristic temperature of $T\sim10\,{\rm keV} \sim 10^8\,$K,
which in combination with the luminosity $L_X\sim10^{37}\,$erg/s gives
an area of $\sim 10^{10}\,$cm$^2$, consistent with the area of a
``polar cap.'' This is the area through which open magnetic field lines pass
for a $R\sim 10\,$km star, 
rotating at $P=1.24\,$ s, with a $B\sim 10^{12}\,$G field.

For the non-pulsating bright X-ray source Sco X-1, $T\sim1\,$keV,
$L\sim10^{38}\,$erg/s, i.e., $R\sim10\,$km directly,
 as expected if the accreting
material is spread over the whole surface.

d) For pulsars, an upper limit to the stellar radius follows from
causality, $\omega R<c$, hence $R<cP/(2\pi)$. For millisecond pulsars,
this gives $R<100\,$km.

e) The moment of inertia of certain pulsars (if they are powered by
rotation) can be measured directly in ``cosmic calorimeters.''
If the luminosity of the Crab nebula ($\approx5\times10^{38}\,$erg/s)
is equated to $I\omega\dot\omega$, for the known period ($P=33\,$ms)
and its derivative of the Crab pulsar (or the known age of the nebula),
the value $I\approx 10^{45}\,$g$\cdot$cm$^2$ is obtained.
A similar, but less secure, argument can be given for the famous eclipsing
pulsar PSR 1957+20
($P=1.6\,$ms, $\dot P\approx10^{-19}$). It is thought that
the power needed to ablate the $0.02M_\odot$ companion is
$\sim10^{38}\,$erg/s (assuming isotropic emission from the pulsar). Again,
$I\sim 10^{45}\,$g$\cdot$cm$^2$ is obtained.

f) Finally, a lower limit to the density can at once be derived for
rotating objects from Newton's formula for keplerian orbital motion:
$\omega_K=\sqrt{GM/R^3}=\sqrt{4\pi\bar\rho/3}$. 
Since $2\pi/P=\omega \le\omega_K$, for any star rotating at a period $P$,
the mean density satisfies
$\bar\rho \ge 3\pi G^{-1}P^{-2}$. With the known value of
Newton's constant, this gives directly
 $\bar\rho >2\times10^{14}\,$g/cm$^3$, for SAX J 1808.4-3658
($P=2.5\,$ms) or the millisecond pulsars, such as PSR 1957+20
($P=1.6\,$ms).

These basic results are subject to many consistency checks, which in
all cases support the basic result that objects with a solid or fluid
surface (i.e., they are not black holes!)
have been identified of dimensions $M\sim 1M_\odot$ and
$R\sim10\,$km:

i) The gravitational energy released in accretion $L\sim GM\dot M/R$
is consistent (for the discussed values $M\sim M_\odot$ and
$R\sim10\,$km) with the mass accretion rate inferred from theoretical
studies of binary evolution.

ii) In some X-ray bursters, the photosphere clearly expands.
Again, spectral fits for the temperature and for the radius of the photosphere
(eq.~[1]), assuming Eddington luminosity, constrain the $M$ -- $R$
relationship, in a manner consistent with the values discussed above.

iii) The surface magnetic field measured from the cyclotron line in X-ray
pulsars agrees, to an order of magnitude ($B_p\sim10^{12\pm1}\,$G), with
the one inferred for radio pulsars, by applying the notion that the spin
down in the latter sources is obtained through balancing the energy loss
in the simple dipole formula $\dot E=-2|\ddot m|^2/(3c^2)$,
where $|m|=B_pR^3/2$, with the kinetic energy loss of a body
of moment of inertia $I= 10^{45}\,$g$\cdot$cm$^2$.

Incidentally, for millisecond pulsars, the value inferred from spin-down,
$B_p\sim 10^{9\pm1}\,$G, is
consistent with the absence of polar cap accretion
 (and of associated pulsations) in X-ray bursters and other LMXBs.
Thus, as far as the magnetic field is concerned, two or three
classes of neutron stars are known---ordinary radio pulsars and
accreting X-ray pulsars ($B\sim10^{12\pm1}\,$G), millisecond
radio pulsars ($B\sim 10^{9\pm1}\,$G), and low-mass X-ray binaries,
 where there is no evidence for such strong magnetic fields
(i.e., $B< 10^{9}\,$G).

iv) The observed long-term spin-up and spin-down of accreting X-ray pulsars
is also consistent with a moment of inertia $I\sim 10^{45}\,$g$\cdot$cm$^2$ ,
for torques which are expected at the mass-accretion rates derived from
the observed X-ray flux, assumed to be $L_X\sim GM\dot M/R\sim0.1\dot Mc^2$,
and the assumption that the lever arm corresponds to an Alfvenic radius,
obtained by balancing the ram pressure with the dipole magnetic pressure,
i.e., $B^2/(8\pi)\sim \rho v_r^2$ at $r=r_A$, 
$\dot M=\epsilon 4\pi r^2\rho v_r$, $B=B_p R^3/r^3$, where $\epsilon\sim1$ is
a geometric factor.

\section{The maximum mass of compact stars}

\subsection{Neutron stars or quark stars?}

It is clear that radio pulsars and some accreting X-ray sources
contain compact objects of properties closely resembling those
known from theoretical models of neutron stars. Specifically,
there can be no doubt that rotating stars of
 $M\sim M_\odot$ and $R\sim10\,$km
exist. However their internal constitution is not yet known.
The expected mass and radius of ``strange'' (quark) stars is similar,
the main difference being in that quark stars of small masses would
have small radii---unlike neutron stars whose radius generally grows
with decreasing mass---(Alcock et al. 1986).
The observed ``neutron stars'' could be made up mostly of neutrons,
but some of them could also be composed partly, or even mostly,
of quark matter.

From the point of view of testing GR, the internal constitution of static
(non-rotating) stars would matter little, as their external metric,
directly accessible to observations, would be independent of their
nature---the only parameter in the unique static, spherically symmetric,
 asymptotically flat solution (the Schwarzschild metric) is the gravitational
mass, $M$, of the central body. However, for rapidly rotating stars, the metric
does vary with  properties of the body other than its mass,
and it would be good to know the precise form of the equation of state
(e.o.s.) of matter at supranuclear density.

As we have seen, at least some low-mass X-ray binaries
(LMXBs) contain stellar remnants of extremely high density,
exceeding $10^{14}\,$g\,cm$^{-3}$, and  many of them are not black
holes because they exhibit X-ray bursts of the type thought to
result from a thermonuclear flash on the surface of an
ultra-compact star. Further, in these long-lived accreting systems
the mass of the compact star is thought to have increased over time
by several tenths of a solar mass above its initial value, and in the process
the stars should have been spun up to short rotational periods. The
compact objects in the persistent LMXBs are expected to be the most
massive stellar remnants other than black holes, hence the most
stringent limits on the e.o.s. of dense matter is expected to be
derived from the mass of the X-ray sources in low-mass X-ray
binaries. Before we discuss how this can be done, let us turn
to the maximum mass.

\subsection{The maximum mass of neutron stars}

One quantity that depends sensitively on the e.o.s. is the maximum mass
of a fluid configuration in hydrostatic equilibrium. For neutron stars
this maximum mass, and in general the mass--radius relationship, is
known from integrating the TOV equations for a wide variety of e.o.s.
(Arnett and Bowers 1977). The mass of rotating configurations is also
known (Cook et al. 1994). Here, I will only briefly review the basic physics
behind the existence of the maximum mass and then give an example 
for strange stars, where the e.o.s. is so simple that the variation of mass
with the parameter describing the interactions can be determined
analytically.

As we know from the work of Chandrasekhar and others, the maximum mass is
reached when the adiabatic index reaches a sufficiently low value that
the star becomes unstable to collapse. In the Newtonian case, this
critical index is $4/3$, corresponding to the extreme relativistic
limit for fermions supplying the degeneracy pressure, when the formula
for kinetic energy of a particle $E=\sqrt{p^2c^2+m_f^2c^4}$ reduces to
$E=pc$.

The very simple argument explaining the instability,
due to Landau, goes like this. There is a balance
between the increasingly negative gravitational binding energy
when a massive sphere of fermions is compressed, and the increasing kinetic
energy of each fermion as it is squeezed into an increasingly
confined volume---each fermion likes to live in phase space of volume
$\sim\hbar$. Of course, as the star is compressed
when its mass is increased,
the fermion momenta increase and the extreme relativistic regime is
aproached, with a corresponding softening of the adiabiatic index.
 The total energy of $N$ particles in a star of mass $M$ bound by gravity,
is up to factors of order unity, 
 $E_{\rm tot}= -GM^2/R + N \bar E$, where $\bar E$ is the mean kinetic energy
of the particles. If the particles are fermions, of mass $m_f$,
their momentum following
from the uncertainty principle is $p=\hbar(N/V)^{1/3}$, and we can take
$V=R^3$ for the volume of the star. In the non-relativistic case,
$E=p^2/(2m_f)$ so $N \bar E=\hbar^2N^{5/3}/(2m_fR^2)$,
and a stable configuration
can be found by minimizing  $E_{\rm tot}$ with respect to $R$.
But in the extreme relativistic case, $E=pc$,  $N \bar E=\hbar cN^{4/3}/R$,
and both terms in $E_{\rm tot}$ are now proportional to $1/R$, so no
minimum energy configuration is found.

In reality, to find the maximum mass configuration,
 one has to solve the TOV equations using a plausible e.o.s.
The TOV equations have essentially the same scaling properties as the
familiar equations of Newtonian hydrostatic equilibrium
$${d\,P\over d\,r}=-{Gm\rho\over r^2},$$
$${d\,m\over d\,r}=4\pi r^2\rho,$$
i.e., if the pressure and density scale with some fiducial density,
$P\propto\rho\propto\rho_0$, then $m\propto r\propto\rho_0^{-1/2}$.
Such scalings allow some general statements to be made about the maximum mass,
such as the Rhoads-Ruffini limit: $M<3M_\odot$, 
if $\rho\ge\rho_0>2\times10^{14}\,$g/cm$^3$.

\subsection{Quark stars}

Conversion of some up and down quarks into strange quarks is
energetically favorable in bulk quark matter (because the Fermi
energy is so high) and it has been suggested that at large
atomic number, matter in its ground state is in the form of
``collapsed nuclei'' with strangeness about equal to the baryon
number (Bodmer 1971). On this assumption,
Witten (1984) discussed the possible
transformation of neutron stars to stars made up of matter
composed of up, down, and strange quarks in equal proportions,
and found the maximum mass of such quark stars as a function
of the density of (self-bound) quark matter at zero
pressure is $\rho_0 \ge 4\times 10^{14}\,$g/cm$^3$.
Detailed models of these ``strange'' stars have been constructed
(Alcock et al. 1986, Haensel et al. 1986). Here, I discuss only the
maximum mass of such stars.

Following Alcock (1991), 
take a gas of any relativistic particles---the e.o.s. is $P_g=\rho_g c^2/3$.
If these are moving in a background of vacuum with uniform energy density
$\rho_vc^2=B$, i.e., negative pressure $p_v=-B$, then the e.o.s. connecting
the total pressure $p=p_g+p_v$, with the total density $\rho=\rho_g +\rho_v$,
is 
$$p=(\rho-\rho_0)c^2/3, \eqno (2)$$
 with $\rho_0c^2=4B$. Witten (1984) showed that
for this simple e.o.s. the maximum mass from the TOV equation is
$M=2M_\odot\sqrt{\rho_1/\rho_0}$, with
$\rho_1\equiv4.2\times10^{14}\,$g/cm$^3$. The scaling $\rho_0^{-1/2}$ is
discussed in the previous subsection.

The physical interpretation of the result is that the relativistic particles
are in fact quarks, and the ``bag constant'' $B$, is a device invented at MIT
to simulate their confinement. The e.o.s.  $p=(\rho-\rho_0)c^2/3$, then,
describes interacting quarks in an approximation to quantum chromodynamics
(QCD) known as the MIT bag model (Farhi and Jaffe 1984). Thus, the maximum
mass found is the maximum mass of static strange (quark) stars.
However, it still depends on the free parameter $\rho_0$.

\subsection{The maximum mass of strange stars}

To illustrate the utility of the scaling law, I will now discuss the
maximum mass of a strange star. First, as already noted by Oppenheimer
and Volkoff (1939), the stellar mass decreases with the fermion mass, so
 to find the maximum mass of
a quark star it is enough to consider massless quarks.
In view of the scaling of TOV equations, the question reduces to that
of finding $\rho_0$, the density of strange matter at zero pressure.
In short,
the maximum mass of a strange star in
the model considered is
 $M_{max}=1.98M_\odot \times(59.8{\rm MeV}/B)^{1/2}$,
 and the least upper bound to the mass of the
strange star is given by the same formula, with $B=B_{min}$, the
lowest possible value of the bag constant. Realistically,
the actual maximum mass of a (non-rotating) strange star will
be smaller by about 10\%
because, because in fact, $m_s>0$.

Currently, the actual value of $B$ cannot be reliably derived from
fits to hadronic masses of the quark-model of nucleons.  Its lowest
possible value can be found by requiring 
that neutrons do not combine to form plasma of deconfined up and
down quarks, or equivalently, that quark matter composed of up and
down quarks in 1:2 ratio is unstable to emission of neutrons
through the reaction $u+2d \rightarrow n$. This implies that the
baryonic
chemical potential at zero pressure of such quark matter satisfies 
(Haensel 1996)
$$\mu_{u,d}(0) > 939.57\,{\rm MeV}. \eqno(3)$$

 As we neglect the masses of up and down quarks in our
considerations, the baryonic chemical potential at pressure $P$ is given by
the expression (Chapline and Nauenberg 1976)
$$\mu(P)=(P+\rho c^2)/n= 4(A/3)^{3/4}(P+B)^{1/4}, \eqno(4)$$ 
where $n$ is the baryon number density, and $\rho c^2=An^{4/3}+B$ is
the energy density. For matter (not in beta equilibrium) composed of
deconfined up and down quarks in 1:2 ratio, $n = n_u =n_d/2$ and
hence  $A = (1+2^{4/3}) (3\hbar c /4) \pi^{2/3} C^{-1/3}$, i.e.,
$\mu(0) \propto (B/C)^{1/4}$, where $C\equiv1-2\alpha_c/\pi$ and
$\alpha_c$ is the QCD coupling constant.
Inequality (1) then becomes
$${B\over C} > 58.9\,\mathrm{MeV}\, \mathrm{fm}^{-3} \equiv B_1.
 \eqno(5)$$
Thus, $B_{min}=(1-2\alpha_c/\pi)B_1$, through lowest order in quark-gluon
coupling. So, for massless interacting quarks, the energy density
at zero pressure is $\rho_0 c^2 =4B \ge (1-2\alpha_c/\pi)\rho_1c^2$.
For massive quarks the expression
for minimum density  becomes more complicated, but we will not need it
to determine  the upper bound to the  mass of a static strange star
in the MIT bag model---it is
enough to consider the e.o.s. of an ulrarelativistic Fermi gas in a
volume with vacuum energy density $B> 0$.

For strange matter in beta equilibrium the number densities  of the
(massless for now) up, down, and strange quarks are equal,  $n_u = n_d
= n_s$, and the energy density is $\rho c^2 = A_s n^{4/3} + B_1$,
with $A_s = 9\hbar c \pi^{2/3} C^{-1/3}/4$, as is appropriate for three
colors per flavour. This gives an equation of state identical to
that of non-interacting quarks, eq. (2),
the only difference being in that the lower bound on
the density at zero pressure, following from conditions of neutron
stability (eqs. [3], [5]), is decreased by the factor $C$ with respect
to the value for an ideal Fermi gas in a bag: 
$$ \rho_0(\alpha_c) =
\left( 1 - {2\alpha_c\over \pi}\right)\rho_0(0). 
$$ 
Thus, through lowest order in the QCD interaction,
the fiducial density is changed, but not the e.o.s.
Since the stellar mass scales as
$\rho_0^{-1/2}$, this implies that the least upper bound on the
mass of the star as a function of the QCD coupling constant is
given for non-rotating strange stars by  
$$  M_{max}(\alpha_c) =
\left( 1-{2\alpha_c\over \pi}\right)^{-1/2} M_{max}(0) \eqno(6) 
$$
through first order in $\alpha_c$. For $\alpha_c =0.6$ this gives a
maximum strange star mass of $ 2.54M_\odot$,
higher  by 27\% than the maximum mass which is obtained for $\alpha=0$.

\section{Measuring the mass of accreting neutron (or strange) stars}

Finally, we have to confront the question how the mass of the compact objects
in LMXBs may be determined. Hopefully, a mass will be measured which will
eliminate a class o equations of state of dense matter.
Unfortunately, application to X-ray bursters
of standard methods for determining the mass function of the
binary---and hence constraining the mass of the compact
X-ray source---is exceedingly difficult,
as the optical emission is usually dominated by that of the
accretion disk (e.g. van Paradijs et al., 1996).
However, reliable mass values obtained by this method
may soon become available, particularly for transient sources,
such as the accreting millisecond pulsar SAX J1808.4--3658.

The mass of the compact object in an X-ray binary may also be
determined by studying the time variability of the radiation flux
formed in the  accretion flow. Specifically, for sufficiently
weakly magnetized stars, a maximum frequency is expected
corresponding to the presence of the innermost (marginally) stable
circular orbit allowed in general relativity 
(Klu\'zniak, Michelson and Wagoner, 1990).
 It has been reported that such a maximum
frequency may have been observed, at least in one system where  quasi
periodic oscillations (QPOs) in the X-ray flux saturate at a
particular value (Zhang et~al. 1998). In this manner, several
e.o.s. were excluded
(Klu\'zniak 1998)
on the understanding that the maximum observed kHz QPO frequency
implies a mass in excess of $2M_\odot$; 
see also Kaaret et al. (1997). 
 Similar considerations (Bulik et~al. 1999)
 exclude static (or slowly rotating) quark stars
  if the minimum density of quark matter is
$\rho_0>4.2\times 10^{14}\,$g/cm$^3$, and the quark matter is taken to be
described by the MIT bag model.

The overall conclusion 
(Klu\'zniak 1998) is that neutron-star matter may be composed
simply of neutrons with some protons, electrons and muons, as models of
more exotic neutron-star matter (including hyperons or pion and kaon
condensates) do
not agree with the simplest interpretation of the kHz QPO data,
namely that the maximum frequency observed in the low-mass X-ray
binary 4U 1820-30, i.e., $1066\,$Hz (Zhang et~al. 1998), is
attained in the marginally stable orbit around a neutron star. 
If the compact stellar remnants in these systems are slowly rotating,
the same conclusion would apply to ultra-dense matter in general,
at densities greater than $4.2\times 10^{14}\,$g/cm$^3$, as 
matter composed of massless quarks would also be excluded for
such densities (Bulik et al. 1999).
 However, as we have seen,
minimum densities smaller than $4.2\times 10^{14}\,$g/cm$^3$
seem possible for more realistic models of
self-bound quark matter, and this would change the conclusion.

For rapidly rotating strange stars the conclusion may be drastically
different, as the metric is greatly modified by a pronounced flattening
of the star (this effect is less important for neutron stars).
In general, the marginally stable orbit is pushed out by this effect,
and a fairly low orbital frequency can be obtained for a low mass star.
This is illustrated in Fig. 1 (taken from Stergioulas et al. 1999)
which exhibits the frequency in the innermost (marginally) stable circular
orbit of general relativity (ISCO) as a function of stellar mass, $M$,
for the Schwarzschild metric [the hyperbola $f_+=2.2\,{\rm kHz}(M_\odot/M)$],
as well as the ISCO frequency for strange stars rotating at Keplerian
frequencies (i.e., maximally rotating, at the equatorial mass-shedding limit),
 for various
values of the density at zero pressure, $\rho_0$ of eq.~(2).
It turns out that for these maximally rotating models, the ISCO is always
at 1.7 to 1.8 km above the stellar surface, the increase of the ISCO
orbital frequency for these models can then be understood in terms of
Kepler's law: $2\pi f\sim\sqrt{G\bar\rho}$, where $\bar\rho$
is the mean density of matter inside the orbit.

\begin{figure}
\includegraphics[width=0.95\columnwidth]{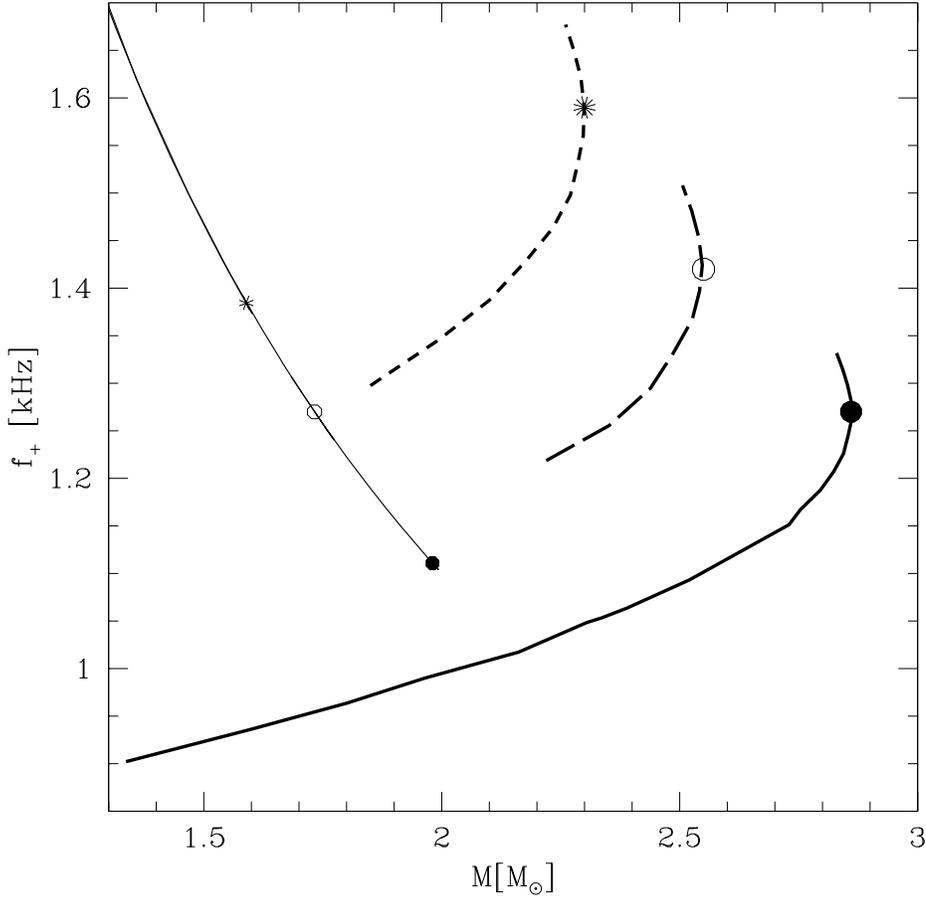}
\caption{The frequency of the co-rotating innermost stable circular orbit
as a function of mass for static models (thin, continuous line)
and for strange stars rotating at the equatorial mass-shedding
limit (thick lines, in the style of Fig. 1). For the static models,
this frequency is given by the keplerian value at $r=6GM/c^2$,
i.e., by
$f_+ = 2198\, {\rm Hz}(M_\odot/M)$, and the minimum ISCO frequency
corresponds to the maximum mass, denoted by a filled circle, an empty circle,
and a star, respectively for
$\rho_0/(10^{14}\,{\rm g\,cm^{-3}})=$ 4.2, 5.3, and 6.5.
Note that the ISCO frequencies for rapidly rotating strange stars
can have much lower values, and $f_+<1\,$kHz can be achieved
for strange stars of fairly modest mass, e.g. $1.4M_\odot$,
if the star rotates close to the equatorial mass-shedding limit.
This figure is from Stergioulas et al. 1999.  
}
\end{figure}

\section{Testing strong-field general relativity with accreting neutron stars}.

There are really two types of objects where strong-field effects of general
relativity are crucial: black holes and accreting neutron (or quark) stars.
Black holes are both attractive and difficult in this context---on the
one hand, their very existence would be impossible in many other theories
of gravity, on the other, their existence is a hypothesis which must
experimentally verified. Possibly this will eventually be achieved
by careful observations of motions in the inner accretion disk
in AGNs and/or black hole binaries.

The existence of neutron stars (or quark stars) would be perfectly
possible in Newtonian gravity (although their detailed properties
would be different from those expected in a general-relativistic world).
But from the point of view of determining the metric, they have the
great advantage, that not only their mass can be measured (as for binary black
holes), but also, at least in some cases, other basic parameters
such as the rotational period and the radius can be determined directly.
Hopefully, this would allow relativistic effects in the accretion flow to
be unambigously resolved.

One class of phenomena which may be helpful in pinning down the external
metric of accreting sources is the relativistic trapping of 
vibrational modes in the inner accretion disk. Indeed, it has been
suggested that the 67 Hz oscillation seen in the source GRS 1915+105
has this origin, and is a signature of the Kerr metric (Nowak et al. 1997).
This is perhaps the most convincing relativistic effect discovered to
date in accreting sources. Unfortunately, the mass of GRS 1915+105
is not known, and there is no independent knowledge of its angular
momentum (the source is a black hole candidate).

Another promising avenue is the search for the marginally stable
orbit (ISCO), expected to exist in accreting neutron stars
(Klu\'zniak and Wagoner 1995) and to show up as a maximum frequency
in the X-ray spectra of LMXBs (Klu\'zniak, Michelson and Wagoner 1990).
Indeed, the recently discovered kHz QPOs in X-ray bursters and other
probable neutron star systems do show some features which are consistent
with their observed frequency being the Keplerian frequency in an accretion
disk terminating close to the marginally stable orbit (Kaaret 1997,
Zhang 1998, Klu\'zniak 1998). But with the data gathered to date, it seems
easier to constrain the e.o.s. of dense matter, on the assumption that
the QPO frequency saturates in the ISCO, than to show that this assumption
is indeed correct. One difficulty is that the physics of accretion disks
is still very poorly understood.

New data is being gathered daily and new
experiments are planned which may lead to a break-through in this field.

I thank the organizers of this School for their wonderful hospitality
in Guanajuato.

\end{document}